\documentclass[preprint]{rmaa}

\usepackage{epstopdf}
\usepackage{textcomp}
\usepackage{multirow} 

\newcommand{\kms}{\mbox{$\,$km s$^{-1}$}}

\newcommand{\subsun}{M$_{\hbox{$\odot$}}$}

\title{Extended Non-Thermal Emission Possibly Associated with Cyg OB2 \#5}

 \author{
 Gisela N. Ortiz-Le\'on\altaffilmark{1},
  Luis F. Rodr\'\i guez\altaffilmark{1,2} and Mauricio Tapia\altaffilmark{3}}
  \altaffiltext{1}{Centro de Radioastronom\'{\i}a y Astrof\'{\i}sica, Universidad
Nacional Aut\'onoma de M\'exico, Campus Morelia} 
\altaffiltext{2}{Astronomy Department, Faculty of Science, King Abdulaziz University, 
P.O. Box 80203, Jeddah 21589, Saudi Arabia}
\altaffiltext{3}{Instituto de Astronom\'\i a, Universidad Nacional Aut\'onoma de M\'exico, Ensenada}

\fulladdresses{
\item Gisela N. Ortiz-Le\'on and Luis F. Rodr\'\i guez: Centro de Radiostronom\'{\i}a
y Astrof\'\i sica, 
UNAM, A. P. 3-72, 
(Xangari), 58089 Morelia, Michoac\'an, M\'exico
(g.ortiz, l.rodriguez@crya.unam.mx).
\item Mauricio Tapia: Instituto de Astronom\'\i a, Universidad Nacional Aut\'onoma de 
M\'exico, Apartado Postal 877, Ensenada, Baja California, CP 22830, M\'exico  
(mt@astrosen.unam.mx).
}

\shortauthor{Ortiz-Le\'on et al.}
\shorttitle{Extended Non-Thermal Emission Near Cyg OB2 \#5}

\SetVolume{50} \SetFirstPage{001} \SetYear{2010}
\ReceivedDate{\today} 
\AcceptedDate{Year Month Day} 

\resumen{
Cyg OB2 \#5 es un sistema binario de contacto (O6.5-7+O5.5-6)
con emisi\'on de radio asociada.
Dos fuentes compactas ($\leq 0\rlap.{''}3$) de radiocontinuo
han sido reportadas anteriormente: la primaria est\'a asociada con la binaria
de contacto y la secundaria es una fuente con forma de arco proyectada a
$\sim 0\rlap.{''}8$ al NE de la primaria. Esta fuente con forma de arco
resulta de la interacci\'on de los vientos de la binaria de contacto 
y de una estrella tipo B en la regi\'on.
En este art\'\i culo reportamos la detecci\'on de una componente extendida
($\sim 30''$) y no-t\'ermica al NE de las componentes compactas.
Proponemos que esta emisi\'on extendida podr\'\i a ser una fuente de fondo
(i. e. una radiogalaxia), emisi\'on gal\'actica extendida, o emisi\'on no-t\'ermica relacionada con
electrones relativistas que se producen en
el choque entre la binaria de contacto y la estrella tipo B
y que son arrastrados a grandes distancias por el viento de la binaria de contacto.
}
 
\abstract{Cyg OB2 \#5 is a contact binary system (O6.5-7+O5.5-6) with associated
radio continuum emission. Two compact ($\leq 0\rlap.{''}3$) radio continuum components
have been reported previously: the primary one is associated
with the contact binary and the secondary one is an arc-like source 
$\sim 0\rlap.{''}8$ to the 
NE of the primary. This arc-like source results from the interaction
of the winds of the contact binary and a B-type star in the region.
In this paper we report the detection of an extended ($\sim 30''$), non-thermal component
to the NE of the compact components. 
We propose that this extended emission could be an unresolved background
source (i. e. a radio galaxy), extended galactic emission,
or non-thermal emission related with relativistic electrons
that are produced in the shock between the contact binary and
the B-type star and that are carried away to large distances by
the wind from the contact binary.
}
      
\keywords{STARS: INDIVIDUAL (CYG OB2 \#5) --- RADIO CONTINUUM: STARS}

\begin{document}

\maketitle

\section{Introduction}

Cyg OB2 \#5 (V729 Cyg, BD +40 4220) is an eclipsing, contact 
binary system (apparent spectral types O6.5-7+O5.5-6; Rauw et al. 1999) 
with a 6.6-day orbital period (Hall 1974; Leung \& Schneider 1978;
Linder et al. 2009). 
As several other luminous O-star systems in the Cyg OB2
association, this source is known to have associated
variable radio emission (e.g. Persi et al. 1985, 1990; Bieging
et al. 1989). 
In addition to the radio emission coincident with the contact
binary, Abbott et al. (1981) and Miralles et al. (1994) reported on the existence 
of a radio ``companion'' 
$0\rlap.{''}8$ to the NE of the contact binary.
Observations by Contreras et al. (1997) revealed that 
this radio source has an elongated shape and lies 
in-between the contact binary and a third star, which was first reported by Herbig 
(1967). Contreras 
et al. (1997) suggested that the proposed NE radio ``companion'' 
actually corresponds to the wind interaction zone between 
the binary system and the tertiary component. Recently, 
Kennedy et al. (2010) reanalyzed all VLA observations 
of Cyg OB2 \#5 and showed that the primary radio source, associated with 
the eclipsing binary,
varies with a period of  $6.7 \pm 0.2$ yr while the flux
from the secondary NE source remains constant
in time. It is now known that the variable radio emission closely
associated with the eclipsing binary comes from
a $\sim$10 milliarcsecond arc-shaped source that traces the wind-collision region 
between the strong wind driven by the contact binary and that of an unseen 
companion (Ortiz-Le\'on et al. 2011).  

\begin{table*}[htbp]
\footnotesize
  \setlength{\tabnotewidth}{2.0\columnwidth} 
  \tablecols{8} 
  \caption{VLA Archive Data Used (Project AR110)}
  \begin{center}
    \begin{tabular}{lccccccc}\hline\hline
    &\multicolumn{3}{c}{6-cm} & & \multicolumn{3}{c}{20-cm} \\
\cline{2-4}
\cline{6-8}
                      & Time on & Flux of & Beam & & Time on & Flux of & Beam   \\
Epoch                 & Source & 2007+404$^a$ & Angular Size$^b$ & & 
Source & 2007+404$^a$ & Angular Size$^b$ \\
& (min)& (Jy)& &  &(min)& (Jy) & \\
\hline
1984 Sep 06  & 30.5   & 4.415$\pm$0.009  & $14\rlap.{''}82 \times 13\rlap.{''}15;~\ $+$47.0^\circ$ &
& 31.0  & 3.977$\pm$0.022 & $50\rlap.{''}08 \times 43\rlap.{''}93;~\ $+$56.5^\circ$  \\
1984 Sep 15  & 31.7   & 4.119$\pm$0.089  & $13\rlap.{''}22 \times 12\rlap.{''}16;~\ $--$49.2^\circ$ &
& 41.7  & 3.618$\pm$0.114 & $45\rlap.{''}96 \times 39\rlap.{''}70;~\ $-$36.2^\circ$  \\
1984 Sep 20  & 32.6   & 4.347$\pm$0.006  & $15\rlap.{''}01 \times 13\rlap.{''}52;~\ $+$64.5^\circ$ &
& 31.8  & 4.011$\pm$0.051 & $52\rlap.{''}91 \times 45\rlap.{''}84;~\ $+$74.7^\circ$  \\
1984 Sep 22  & 17.5   & 4.151$\pm$0.015  & $13\rlap.{''}92 \times 12\rlap.{''}43;~\ $+$51.7^\circ$ &
& 17.0  & 3.856$\pm$0.052 & $48\rlap.{''}10 \times 43\rlap.{''}17;~\ $+$51.6^\circ$  \\
1984 Sep 24  & 45.5   & $4.372\pm$0.016  & $14\rlap.{''}22 \times 12\rlap.{''}81;~\ $--$70.2^\circ$ &
& 45.0  & 4.132$\pm$0.025 & $46\rlap.{''}68 \times 42\rlap.{''}19;~\ $-$55.8^\circ$  \\
1984 Sep 28  & 40.0   & 4.211$\pm$0.009  & $13\rlap.{''}28 \times 12\rlap.{''}33;~\ $--$0.71^\circ$ &
& 40.5  & 4.015$\pm$0.033 & $44\rlap.{''}83 \times 40\rlap.{''}53;~\ $+$2.55^\circ$  \\
\hline\hline
\tabnotetext{a}{The phase calibrator for all the observations was 2007+404.}
\tabnotetext{b}{Major axis $\times$ minor axis; position angle, for a 
\sl (u,v) \rm weighting of ROBUST = 0 (Briggs 1995).}
    \label{tab:1}
    \end{tabular}
  \end{center}
\end{table*}

\begin{figure*}
\centering
\includegraphics[scale=0.3, angle=0]{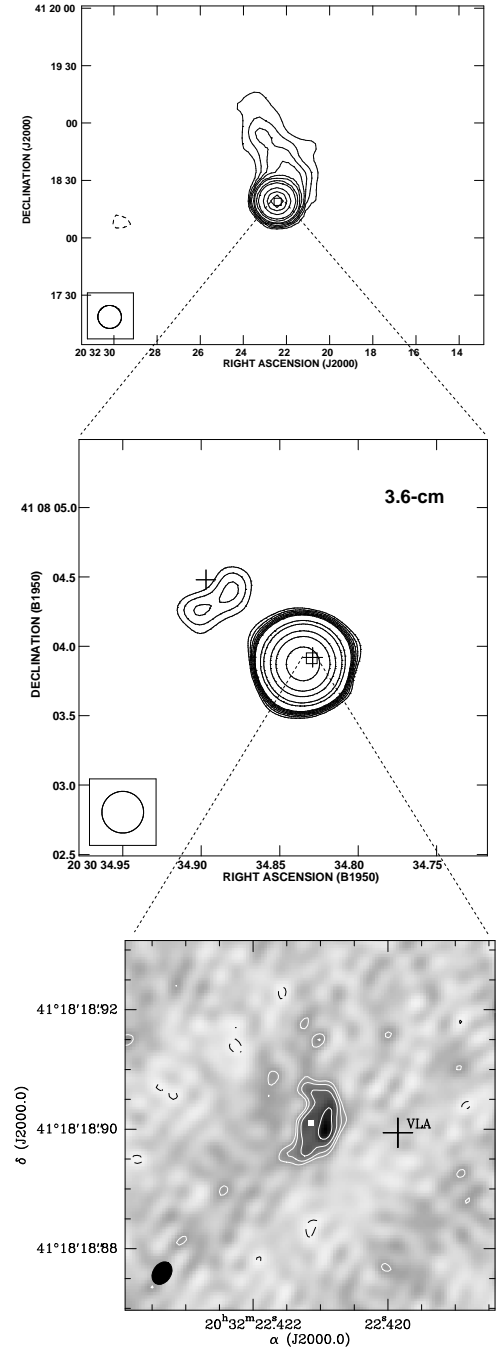}
 \caption{(Top) VLA contour image of the 6-cm continuum emission toward
Cyg OB2 \#5. Contours are -3, 3, 4, 5, 6, 8, 10, 15, 20, 40, 60 and 80
times 79 $\mu$Jy, the rms noise of the image. 
The synthesized beam, shown in the bottom left corner,
has half power full width dimensions of
$12\rlap'{''}3 \times 12\rlap.{''}0$, 
with the major axis at a position angle of $+13^\circ$. 
The bright source at the center coincides with Cyg OB2 \#5 and comprises
the emission from the contact binary and the compact NE component. The
extended emission to the NE is first reported here. (Middle) High-angular resolution
3.6-cm emission from Cyg OB2 \#5, from Contreras et al. (1997). The source at the center
is the emission from the contact binary and that to the NE corresponds to the compact NE
component. The crosses mark the position of the contact binary (center) and of the
known B-type star to the NE (Herbig 1967). (Bottom) VLBA image at 3.6 cm from Ortiz-Le\'on
et al. (2011). The arc-shaped source traces the wind-collision region between
the wind of the contact binary (whose position is indicated with a cross)
and the undetected companion (whose estimated position is indicated with
a white square).
}
  \label{fig1}
\end{figure*}

\begin{figure*}
\centering
\includegraphics[scale=0.45, angle=0]{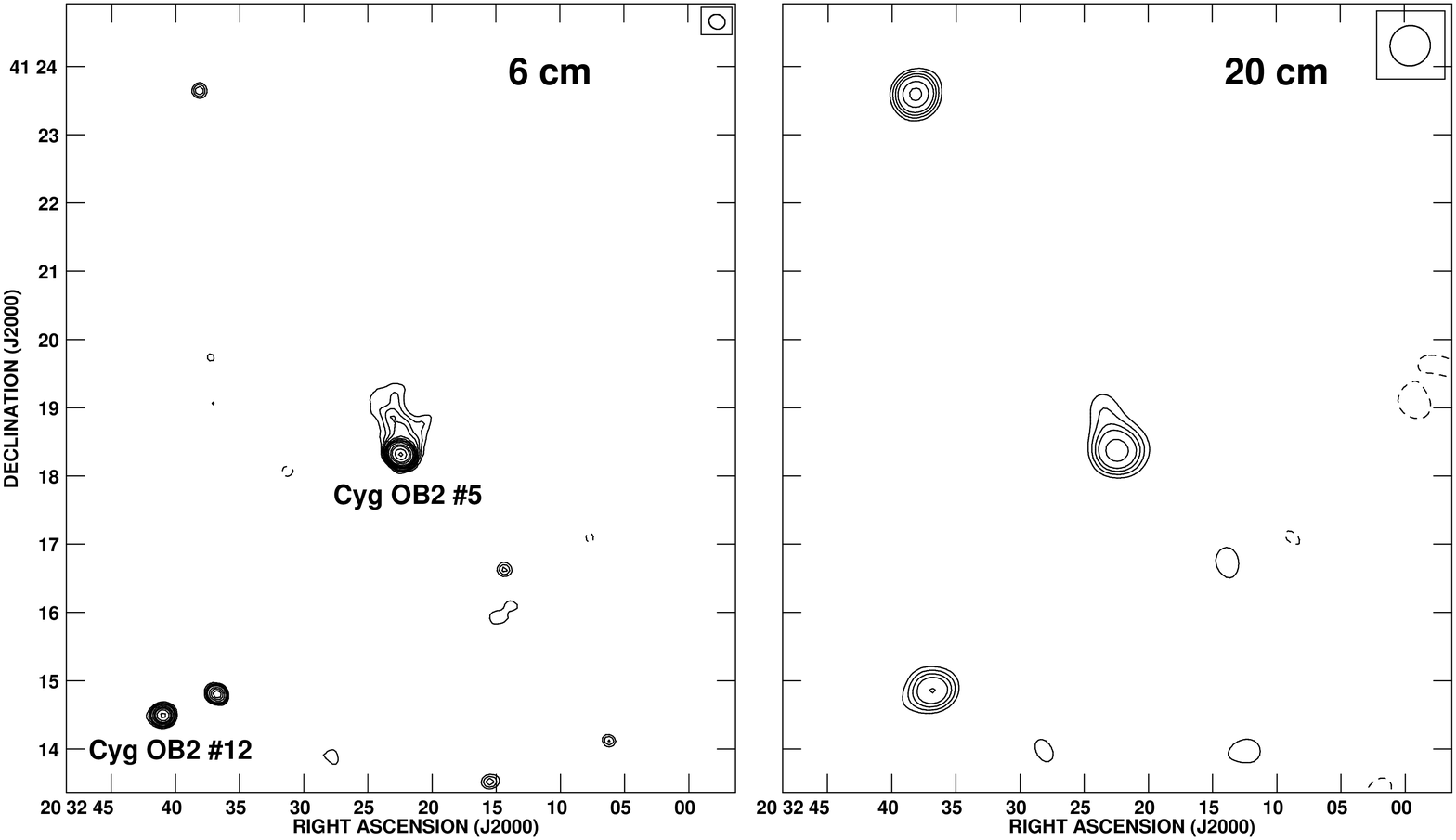}
 \caption{(Left) A larger VLA contour image of the 6-cm continuum emission toward
 Cyg OB2 \#5. The stars Cyg OB2 \#5 and Cyg OB2 \#12
 are identified. The other radio sources do not have known counterparts.
 (Right) VLA contour image 
 of the 20-cm continuum emission toward
 Cyg OB2 \#5. Contours are -3, 3, 4, 5, 6, and 8
 times 725 $\mu$Jy, the rms noise of the image.
 The synthesized beam, shown in the top right corner,
 has half power full width dimensions of
 $35\rlap'{''}8 \times 35\rlap.{''}1$,
 with the major axis at a position angle of $-44^\circ$.
These images are not corrected for the
 primary beam response. Several of the sources detected in the 6 cm
 image, including Cyg OB2 \#12, are not detected in the 20 cm image given the much larger noise
 of the latter.}
   \label{fig2}
   \end{figure*}

As part of a VLA archive analysis of this source, we concatenated
6 and 20-cm data obtained in the compact configuration D in six epochs
during September of 1984. These deep images reveal the presence
of an extended emission component apparently associated
with Cyg OB2 \#5 whose characteristics and possible nature
are discussed here.

\section{Data Reduction}

The archive data from the Very
Large Array (VLA)
of the NRAO\footnote{The National Radio
Astronomy Observatory is operated by Associated Universities
Inc. under cooperative agreement with the National Science Foundation.}
were edited and calibrated using the software package Astronomical Image
Processing System (AIPS) of NRAO. The parameters of
the observations are given in Table 1. The absolute amplitude
calibrator for five of the six epochs was 1331+305, with an
adopted flux density of 7.49 Jy at 6 cm and 14.39 Jy at
20 cm. The amplitude calibrator for
1984 September 22 was 0137+331, with an
adopted flux density of 5.43 Jy at 6 cm and 15.29 Jy at
20 cm. The phase calibrator was 2007+404 for all six epochs, with
the bootstrapped flux densities given in Table 1.

An image of the 6 cm emission made with the ROBUST parameter of AIPS
set equal to 0 is shown in the top part of Figure \ref{fig1}.
This image was made using only baselines longer that 1 k$\lambda$ to suppress
structures larger that $\sim3\rlap.{'}4$.
The image shows a bright source at the center. This emission is the spatially
unresolved combination of the emission from the contact binary and the
NE component. In addition to this bright source there is an extended,
faint component to the NE. 
This component is not reported previously, most probably because
for the VLA configurations A and B, structures as large as this
component are resolved out. At C and D configurations, the archive data analyzed by us
is much better and sensitive than other databases available. We discuss its possible nature
in what follows. 

The 6 cm and 20 cm data were self-calibrated in phase using all
the positive clean components of 
the sources in the field shown in Figure 2 to model the emission.
In this Figure we show a larger part of the same image, where
it can be appreciated that the
only significant features (above 4-sigma) are Cyg OB2 \#5, the extension to the 
NE and several compact sources in the region. In this same Figure we show a 20 cm image
of the same region. To obtain a reliable comparison, both the 6 and 20-cm data
were restricted to baselines larger than 1 k$\lambda$.
The 20 cm data are difficult to use because of contamination
from extended emission, and the field
contains other features of strength comparable to that proposed to
be associated with Cyg OB2 \#5. However, we believe the reality
of the NE feature because it is present at both wavelenghts.

\section{Interpretation}

\begin{figure}
\centering
\includegraphics[scale=0.35, angle=0]{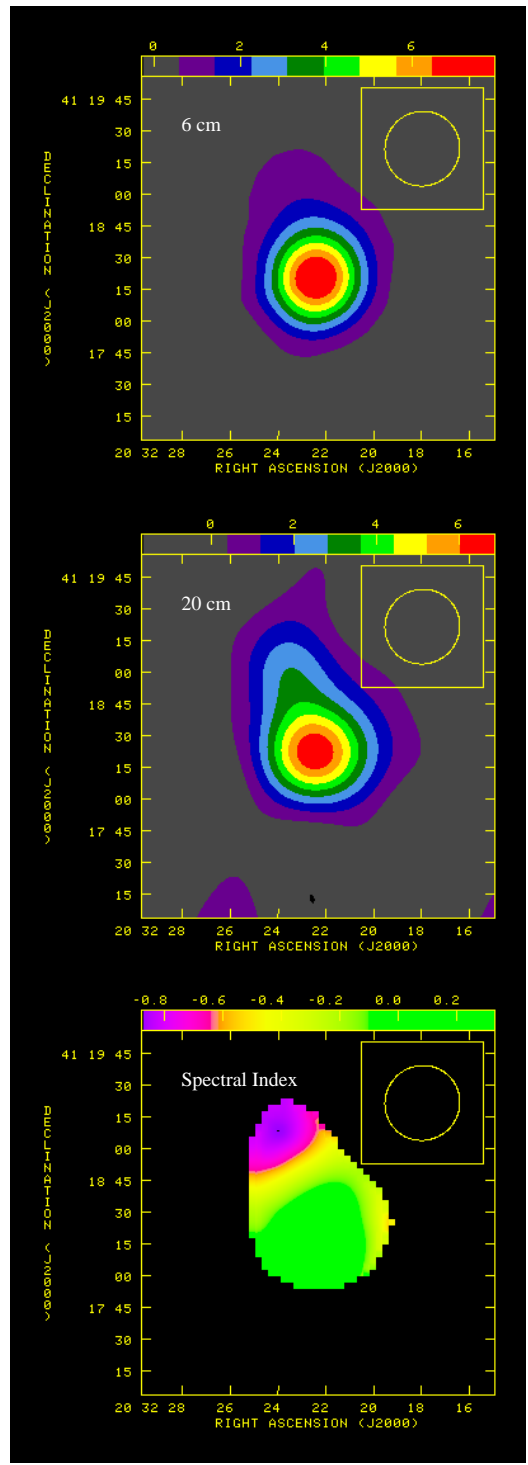}
 \caption{VLA color images of the 6-cm (top) and 20-cm (middle)
continuum emission toward
 Cyg OB2 \#5.
The spectral index of the emission is shown in
 the bottom panel. 
 The flux density and spectral index scales are given in the upper part
  of the panels.
 The restoring beam, shown in the top right corner,
 has half power full width dimensions of
 $36{''} \times 35{''}$,
 with the major axis at a position angle of $-44^\circ$.
 }
   \label{fig3}
   \end{figure}

In order to investigate the character of the extended emission, we determined 
the spectral index of the bright source and the faint component from the 6
and 20-cm data. The 6-cm
data was degraded in angular resolution to match that of the
20-cm data. In Figure \ref{fig3} we show images of radio continuum at
6 and 20 cm in the upper and middle panels, respectively. 
A spectral index map was made with these images using the AIPS task COMB and 
it is shown at the bottom panel of Figure \ref{fig3}. 

From analysis of the images shown in Figure \ref{fig3}, we determine
for the bright compact component a flux density
of $7.9 \pm 0.2$ mJy at 6 cm and $6.8 \pm 0.4$ 
mJy at 20 cm. For the extended emission we determine a flux density
of $2.1 \pm 0.4$ mJy 
at 6 cm and $4.1 \pm 1.0$ mJy  at 20 cm. The spectral index is $0.12\pm0.05$ 
for the bright compact component, consistent with the value expected for the high state
of the variable flux, while for the extended component this value
is $-0.6\pm0.3$, 
indicating that the faint emission is non-thermal, most likely
of a synchrotron nature. The spectral index of the extended component
is consistent with the value of $-0.5$ expected for synchrotron
emission from shock accelerated electrons (Jun \& Jones 1999).
We then propose that the extended emission to the NE of the 
bright component could be i) an unresolved background source 
(i. e. a radio galaxy), ii) extended galactic emission, or 
iii) non-thermal emission of relativistic 
electrons that are produced in the wind-collision region between the 
stellar winds from the contact binary and the B-type star and 
that are carried away to large distances by the wind from 
the contact binary.

\subsection{A Background Source?}

Using the formulation of Fomalont et al. (1991) we estimate that the 
probability of finding a background source with a flux of $2.1$ 
mJy at 6 cm in a  $2'\times2'$ box is only 1.1\%. This probability is 
small, however  we cannot rule out the possibility that the extended emission 
can be a background source projected near of Cyg OB2 \#5. 
An image with better angular resolution than that of 
Figure \ref{fig1} could reveal the morphology of a radio galaxy
and favor this interpretation.
Lacking at present such a radio image, we searched through the available infrared 
survey-images in order to look for any possible infrared counterpart to the extended 
non-thermal radio source reported here that may give an indication of its nature.

\begin{figure*}
\centering
\includegraphics[scale=0.5]{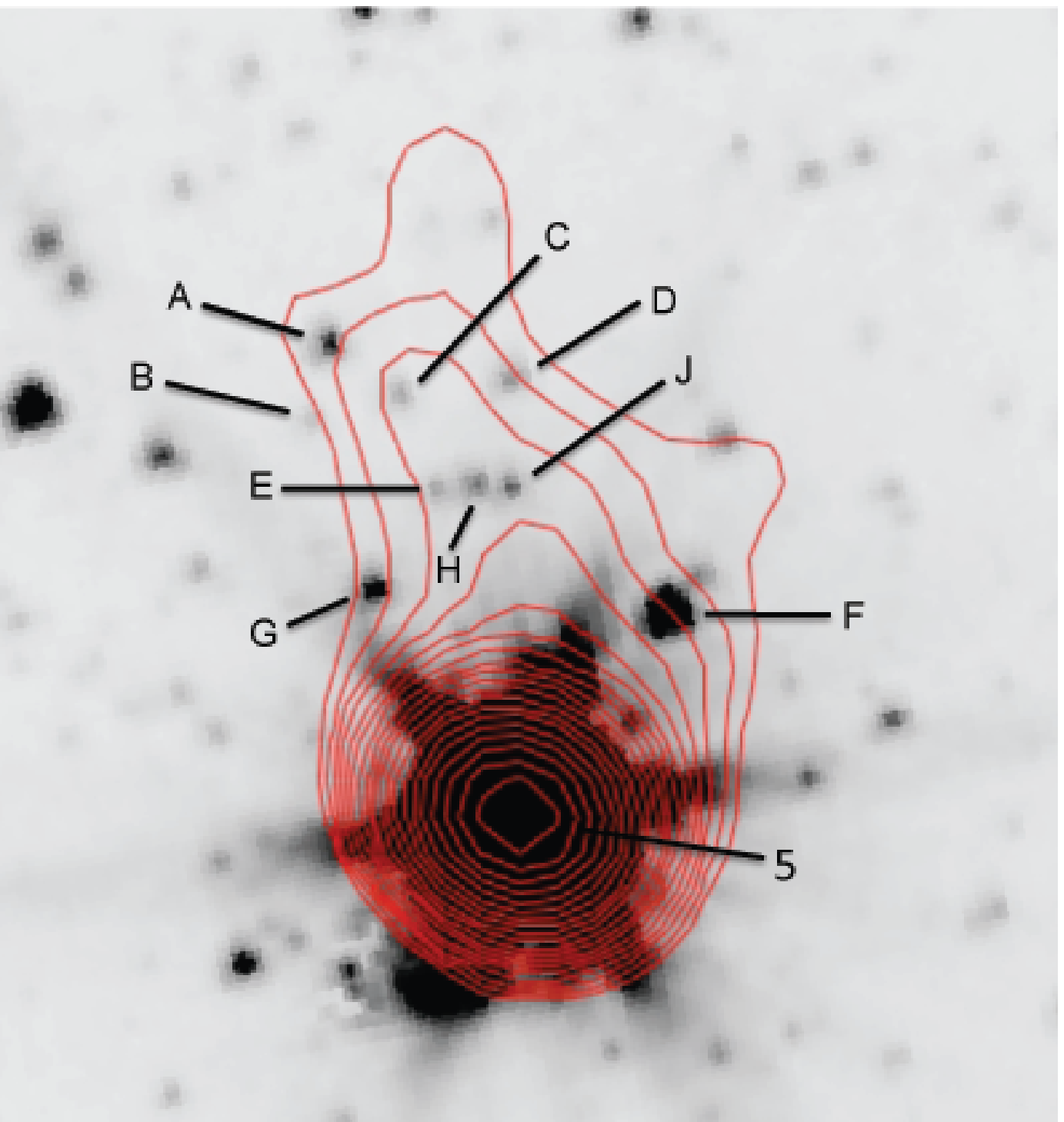}
 \caption{Identification chart of all point sources listed in Table 2 over an IRAC 4.5 $\mu$m
 image. 
 All sources lie within or close to the (shown) contours of the extended faint 6-cm 
 emission to the north of Cyg~OB2~\#5. The field of view is $86 \times 93$ square arcsecs. 
 North is to the top  and east to the left.}
   \label{fig4}
   \end{figure*}

We analyzed the available 2MASS images in the $JHK$ bands as well as those taken
by the {\it Spitzer Space Observatory} at 3.6, 4.5, 5.8 and 8 $\mu$m (IRAC) and at 
24~$\mu$m (MIPS). Only unresolved sources were found within or close to the radio contours on 
all these images. In Figure \ref{fig4}, an identification chart is presented on a 4.5 $\mu$m 
gray-scale
IRAC frame. Aperture photometry with appropriate ``sky'' subtraction of all sources was 
performed on the Spitzer images and near-infrared photometric data was retrieved from the 
2MASS Point Source Catalogue. When possible, these were supplemented by differential aperture 
photometry on the $JHK$ images for those sources omitted in the published catalog.
Nevertheless, diffraction spikes prevented us from obtaining reliable near-infrared 
photometry for sources D, J and H. The near-infrared photometry of Cyg~OB2~\#5 is from 
Torres-Dodgen et al. (1991) 

The results are given in Table 2. When the photometric uncertainties were of the order 
of 10 to 15\%, the results are given with one decimal; otherwise, these are around 5\%.
The data for the bright binary star Cyg~OB2 \#5 are also included for comparison.
From their location on all infrared two-color diagrams, we could assert that sources
B, C, 
E, F and, naturally, 
Cyg~OB2~\#5 have 1.2 to 8 $\mu$m colors totally consistent with being reddened 
mid- to early-spectral type photospheres. Sources A, H and J display 
1 to 24 $\mu$m colors consistent with Class~II pre-main sequence stars. Sources D and
G also show slight infrared excesses. With the probable exception of sources B and E, 
the characteristics of this small group of infrared point sources is representative of 
the population of young stellar objects in Cyg~OB2 (cf. Hora et al. 2011).

Still, the above exercise does not prove that any of the infrared-excess sources cannot 
be an active radio galaxy, since some of these may have similar infrared colors (e.g.
Donley et al. 2008, Kouzuma \& Yamaoaka 2010). Nevertheless, we consider this to be very 
unlikely, as further supported by the results of the complete Spitzer Legacy Survey of the 
Cygnus-X region described by Hora et al. (2011). These authors found that the known AGN
galaxies detected by IRAC in the region were all fainter than [4.5]~$\simeq~14$, and 
were considerably redder than the sources reported here.

\begin{table*}[htbp]
\footnotesize
  \setlength{\tabnotewidth}{2.0\columnwidth} 
  \tablecols{10} 
  \caption{Coordinates and IR magnitudes of sources close to the extended radio source}

  \begin{center}
    \begin{tabular}{ccclllllll}\hline\hline
 & \multicolumn{2}{c}{(J2000)} & \sl J & \sl H & \sl K &  [3.6] & [4.5] & [5.8] & [8] \\
\hline
A&  20 32 23.8 & 41 18 57  & $<16$ &15.8  & 14.9   &13.30  &12.66  &12.64  &12.1\\
B&  20 32 23.9 & 41 18 51  & $<16$ &15.64 & 14.84  &14.69  &14.73  &15.2   & -  \\
C&  20 32 23.3 & 41 18 53  &15.38  &14.56 & 14.15  & 14.11 &14.14  &14.6   &14.7\\ 
D&  20 32 22.5 & 41 18 54  &  -    & -    & -      & 13.63 &13.51  &14.3   &14.1\\
E&  20 32 23.0 & 41 18 45  &16.1   &14.62 & 14.15  & 14.0  &14.2   &14.7   &   -\\ 
H&  20 32 22.7 & 41 18 46  &  -    &  -   &  -     & 13.5  &13.3   &13.7   &13.2\\
J&  20 32 22.5 & 41 18 46  &  -    & -    & -      &13.6   &13.6   &13.9   &13.2\\  
G&  20 32 23.5 & 41 18 37  &15.52  &14.52 & 13.77  &12.70  &12.42  &12.45  &12.5\\
F&  20 32 21.3 & 41 18 35  &11.39  &10.94 & 10.67  &10.83  &10.80  &11.07  &11.4\\ 
5&  20 32 22.4 & 41 18 19  & \hspace{1.5pt} 5.24  & \hspace{1.5pt} 4.55 & \hspace{1.5pt} 4.27   & \hspace{1.5pt} 3.5    &\hspace{1.5pt} 3.6    & \hspace{1.5pt} 3.7  & \hspace{1.5pt} 3.8 \\
\hline\hline
    \label{tab:2}
    \end{tabular}
  \end{center}
\end{table*}

\subsection{Extended Galactic Emission?}

Cyg OB2 \#5 is in the local (Cygnus) arm of the Galaxy, very close to
the galactic plane and
surrounded by complex, extended low surface brightness emission, making
this possibility likely. However, the
image shown in Figure 2 shows that the only clear extended component is
the one associated with Cyg OB2 \#5.

\subsection{Relativistic Electrons from the Shock?}

The third possibility for the explanation of the extended emission
is interesting because it would imply that small systems of 
massive stars with non-thermal emission like Cyg OB2 \#5  can be
sources of relativistic electrons for the interstellar medium.
In this explanation the relativistic electrons produced in the interaction
between the winds of the contact binary and the B-type stellar companion
(that produce the non-thermal NE compact component) could be carried
away, together with magnetic fields, by the wind of the contact binary
to produce faint, extended non-thermal emission.

We then estimate the minimum energy of the relativistic electrons  
accelerated in the wind interaction region
and the magnetic field
that produce the non-thermal secondary component $0\rlap.{''}8$ to the NE of the 
primary and then ask if these parameters are consistent with
the electrons being able to travel to a distance 
equal to the size of the extended non-thermal emission.
This estimate is relevant to check that the relativistic electrons
produced in the secondary compact
NE region can travel into the interstellar medium
and produce the extended component.

Assuming minimum total (magnetic plus
relativistic electrons) energy
and a homogeneous spherical source, we can
use the equations of Pacholczyk (1970) to roughly estimate the magnetic
field and the relativistic electron minimum energy in the secondary NE component.
In the absence of a detailed
knowledge of the nature of the source, these assumptions are consistent 
with the available data. We adopt
a flux of 0.8 mJy at 6 cm for the region (Kennedy et al. 2010) and 1.7 kpc as
the distance to the source (Torres-Dodgen et al., 1991). The radius of the
emitting region is $\sim 2.5\times 10^{15}$ cm. We also use 1$-$30 GHz as the frequency range
observed. This gives a magnetic field of 2.0 mGauss and an electron minimum
energy of $8\times10^{39}$ ergs.

The Lorentz $\gamma$ factor for these relativistic electrons is $2\times10^3$ and the 
lifetime of them in such a magnetic field is $5\times10^3$ years (Pacholczyk 1970). Hence, if the 
electrons are  carried away by the wind from the contact binary, whose terminal 
velocity is adopted to be  \mbox{$\sim 2200$ \kms} (Bieging et al. 1989), then they 
can travel emitting centimeter synchrotron radiation up to a distance of $\sim 10$ pc.
This distance corresponds to an angular size of $\sim 20'$ which 
is much greater than the size of the extended emission, of only $\sim 30''$.
However, the post-shock velocity of
the wind will be much smaller than the original wind velocity (1/4 for the case of
an adiabatic shock) and the distance that the relativistic electrons can
travel will be correspondingly smaller ($\sim 5'$) but still much larger
that the size of the extended emission ($\sim 30''$).
Additionally, the obliquity of the shock geometry can further reduce 
the post-shock velocity of the wind, but we have a margin of a
factor of $\sim$10 and conclude that the size 
of the extended emission detected is not an obstacle for associating it with 
relativistic electrons produced in the compact
secondary NE source. 
We have also considered the relevance of inverse Compton (IC) cooling. Even when this cooling
is more important than synchrotron cooling close to the star, the gas that carries
the relativistic electrons is moving away very fast from the star and
it can be shown (see Appendix) that IC cooling will not produce a major energy loss.

We now discuss 
if the electron energy of \mbox{$8\times10^{39}$} ergs in the secondary NE source  
can be produced by transforming a fraction of the kinetic energy of the binary wind into 
the acceleration of the electrons. The secondary NE source subtends a full
opening angle of 
$\sim 30\!\!\!\phantom{a}^{\circ}$, as measured from the primary component and this suggest 
that it intercepts about $2\%$ of the contact binary wind (see middle panel
of Figure 1). Adopting a mass-loss 
rate of \mbox{$5\times10^{-5}$ \subsun \, yr$^{-1}$}  (Contreras et al. 1996) 
and a terminal velocity of \mbox{$2200$ \kms} (Bieging et al. 1989)
for the contact binary wind we obtain 
a total wind kinetic power of \mbox{$7.8\times10^{37}$ ergs s$^{-1}$}. 
The secondary NE source width  
(over the line that joins it with the binary) is $\sim 0\rlap.{''}1$.
This corresponds to a distance of \mbox{$2.6\times10^{15}$ cm}, that the wind would 
go over in \mbox{$1.2\times10^7$ s} (0.4 years). Then, during this time the contact 
binary wind deposits into the secondary compact NE source an energy of
\mbox{$7.8\times10^{37}$ ergs s$^{-1} \times 1.2\times10^7$ s $\times 0.02$} 
\mbox{$=1.8\times10^{43}$ ergs}. Therefore, to explain the energy of the
relativistic electrons in the secondary compact NE source ($8\times10^{39}$ ergs)
produced by the shock 
of the stellar winds from the contact binary and the B-type star,  
we need an acceleration mechanism
that converts $0.04\%$ of the wind's kinetic energy from the contact binary into
the acceleration of the relativistic electrons.
We note that this relatively low efficiency is
consistent with the results of studies (Eichler \& Usov 1993; 
Ellison \& Reynolds 1991; Blandford 
\& Eichler 1987) that indicate that 
the relativistic electron energy is unlikely to be 
more than 5\% of the total available 
shock energy, and could conceivably be much less.
For the non-thermal extended region we use a
flux density of 2.1 mJy at 6 cm and an angular size of $30''$,
with the source located at a distance of 1.7 kpc from the Sun. 
This gives a magnetic field of 
\mbox{$\sim 28 \, \mu$Gauss} and an electron minimum energy of 
\mbox{$\sim 1\times 10^{43}$ ergs}.

An independent consistency check between the synchrotron luminosity and the
luminosity of the WCR can be made following
Kennedy et al. (2010). For the luminosity of the wind-collision-region we
obtain $L_{WCR} \simeq 1.6 \times 10^{36}$ erg s$^{-1}$. The radio synchrotron 
luminosity $L_{syn}$ arising from the WCR, can be estimated from
$L_{syn} \simeq  10^{-8}~L_{WCR}$ (Chen \& White 1994; Pittard \& Dougherty 2006), giving an expected
synchrotron luminosity from the WCR of $1.6 \times 10^{28}$ erg s$^{-1}$.
Integrating the radio luminosity in the 1 to 30 GHz range, we obtain a synchrotron luminosity of
$5.4 \times 10^{28}$ erg s$^{-1}$, that given the uncertainties is consistent
with the expected value.
The radio luminosity of the extended NE component is about three times larger
than that of the compact NE component. Since under our assumptions these radio
luminosities are coming from the kinetic power of the wind, this implies that
a few more times power is dissipated in the extended component as compared to the
compact component.

\subsection{Timescale for the acceleration of the relativistic electrons}

Finally, we estimate the characteristic timescale for acceleration of the relativistic electrons
in the interaction region
of the stellar winds from the contact binary and the B-type star, that is,
the compact secondary NE source. This is an important estimate to make since, as
shown before, the crossing time of the contact binary wind across 
this region is of only \mbox{$1.2\times10^7$ s} (0.4 years).
Following Loeb \& Waxman (2000), this characteristic timescale, $t_{acc}$,
is given by

$$\biggl[{{t_{acc}} \over {hr}} \biggr] \sim 1.4 \biggl[{{\gamma} \over {10^3}} \biggr]
\biggl[{{B} \over {mG}} \biggr]^{-1}
\biggl[{{v_{sh}} \over {10^3~km~s^{-1}}} \biggr]^{-2},$$

\noindent where $\gamma$ is the Lorentz factor of the relativistic electrons,
$B$ is the magnetic field of the region, and $v_{sh}$ is the shock velocity.
Adopting $\gamma = 2 \times 10^3$, $B = 2.0$ mGauss, and $v_{sh} =$ 2200 km s$^{-1}$,
we obtain $t_{acc} \sim 0.3$ hours. Then, the acceleration takes place very
quickly, on a timescale much shorter than the wind crossing time.

\section{Conclusions}

We present the analysis of VLA archive data of the Cyg OB2 \#5 system
taken during 1984 in the low angular resolution D configuration.

Concatenating 6 and 20-cm data, we detect an extended region of 
non-thermal emission to the NE of the compact components of Cyg OB2 \#5.
This faint emission could be an
unresolved background source,
like a radio galaxy, but photometry performed on archival
2MASS and {\it Spitzer} images on all infrared sources in the vicinity suggests that this 
possibility is highly unlikely. It is also possible that the
emission is produced by
synchrotron radiation of the relativistic electrons produced in the shock 
of the stellar winds from the contact binary and the B-type star to its NE
and carried away at large distances by the stellar wind of the
contact binary. However, we cannot rule out the possibility that the emission
arises in an extended galactic feature unrelated to the stellar system.

We end by noting that Cyg OB2 \#5 is a remarkable system in that it shows radio
structures over different scales, from about 10 milliarcseconds
to about 30 arcseconds, a range of $3 \times 10^3$
(see Fig. 1). At the 10 milliarcsecond scale, Ortiz-Le\'on
et al. (2011) have used Very Long Baseline Array 
observations to detect the non-thermal, arc-like structure that traces
the wind-collision region betwen the wind of the contact binary and that of 
an unseen nearby companion. At the scale of $\sim0\rlap.{''}1$, one can detect the
thermal (free-free) emission from the wind of the contact binary (Rodr\'\i guez et al.
2010). About $1''$ to the NE of the contact binary, one finds the nonthermal
compact NE component (Contreras et al. 1999) that results from the interaction of
the winds of the contact binary and that of a known B-type star
(Herbig 1967). Finally, in this paper we propose that an extended
($\sim30''$) structure detected to the NE could also be associated with this multiple
stellar system.

\acknowledgments
We thank an anonymous referee for suggestions that greatly improved
our original Figures 1 and 2.
GNOL and LFR are thankful for the support
of DGAPA, UNAM, and of CONACyT (M\'exico).
MT acknowledges DGAPA/PAPIIT grant No. 100210.
This research has made use of the SIMBAD database, 
operated at CDS, Strasbourg, France.
This work makes use of archival data obtained with the Spitzer Space Telescope, which is operated 
by the Jet Propulsion Laboratory, California Institute of Technology under a contract with NASA. 
Support for this work was provided by an award issued by JPL/Caltech. Also of data products from 
the Two Micron All Sky Survey, which is a joint project of the University of Massachusetts and the 
Infrared Processing and Analysis Center/California Institute of Technology, funded by the National 
Aeronautics and Space Administration and the National Science Foundation. 

\vskip0.5cm

\centerline{APPENDIX}

\begin{appendix}

\section{Inverse Compton cooling in gas moving away from a star}

The rate of inverse Compton energy loss of a relativistic electron at a distance $r_0$
of a star with luminosity $L$ is given by (Pittard et al. 2006):

$${{d \gamma} \over {d t}} = - {{\sigma_T \gamma^2 L} \over {3 \pi m_e c^2 r_0^2}},$$

\noindent where $\gamma$ is the Lorentz factor of the electron, $\sigma_T$ is 
the Thomson cross section,
$m_e$ is the electron mass, and $c$ is the speed of light.

Integrating this equation, we find that the timescale for a relativistic electron to
lose one half of its initial energy is given by

$$t_S = {{3 \pi m_e c^2 r_0^2} \over {\gamma_0 \sigma_T L}}.$$

However, in the scenario discussed in the paper, the gas is rapidly moving away
from the star. In this case, the energy loss equation can be put as:

$${{d \gamma} \over {d t}} = - {{\sigma_T \gamma^2 L} \over {3 \pi m_e c^2 r^2}},$$

\noindent where the radius $r$ is increasing with time and is given
by $r = r_0 + v t$, where $r_0$ is the initial radius considered
and $v$ is the wind velocity.
On the other hand, the crossing time of the gas
from the star to the shock region is given by $t_C = r_0/v$.

Solving the previous equation it can be found that for the relativistic electron to
reach infinity without having lost one half of its energy the following condition
should be fullfilled:

$$t_C \leq t_S,$$

\noindent while for this loss to take place in finite distances we need

$$t_C~ \textgreater~ t_S.$$

In the case discussed in the paper, we have $\gamma_0 = 2\times10^3$, $r_0 = 2.0 \times 10^{16}$ cm,
$L = 10^6~L_\odot$ (Linder et al. 2009) and $v = 2200$ km s$^{-1}$.
However, the luminosity of Linder et al. (2009) was estimated for a distance of
925 pc. Correcting for the distance of 1.7 kpc used in our other estimates,
we obtain $L = 3 \times 10^6~L_\odot$.
We then obtain $t_S \simeq 6$ yr and $t_C \simeq 3$ yr. We are then in the
case $t_C \leq t_S$ and inverse Compton scattering produced
by the radiation field of the contact binary is not expected to dominate the
energy loss. 

The B-type star to the NE of the contact binary will have a much smaller luminosity
but it will produce important inverse Compton losses
to the relativistic electrons that pass very close to it. To estimate its
effect on the bulk of the relativistic electrons, we note that the angular radius 
of the compact NE source ($2.5 \times 10^{15}$ cm) is about an order of magnitude
smaller than the distance between the contact binary and the NE compact source
($2.0 \times 10^{16}$ cm). We also note that its approximate spectral type (B0 V-B2 V)
was derived by Contreras et al.(1997) from Hipparcos photometry. Adopting the mean
spectral type of B1 V, we estimate a bolometric luminosity
of $L \simeq 10^4~L_\odot$ (Panagia 1973), about 300 times less than that of
the contact binary.
Since the effect of the inverse Compton scattering
goes as luminosity over distance squared, we conclude that, on the bulk, the
effect of the B-type star will be a few times smaller than that of
the contact binary.

\end{appendix}


\end{document}